\newcommand{\beq}{\begin{equation}}
\newcommand{\eeq}{\end{equation}}
\newcommand{\beqa}{\begin{eqnarray}}
\newcommand{\eeqa}{\end{eqnarray}}
\newcommand{\ket}[1]{| #1 \rangle}

\newcommand{\opa}{\hat{a}}

\documentclass[twocolumn,showpacs]{revtex4}

\begin{document}


\title{Sub-wavelength lithography over extended areas}

\author{Gunnar Bj\"{o}rk}
\thanks{On leave from Department of Electronics, 
Royal Institute of Technology (KTH), Electrum 229, 
SE-164 40 Kista, Sweden}
\email{gunnarb@ele.kth.se}
\homepage{http://www.ele.kth.se/QEO/}
\affiliation{Departamento de \'{O}ptica, 
Facultad de Ciencias F\'{\i}sicas, 
Universidad Complutense, 28040 Madrid, Spain}

\author{Luis L. S\'{a}nchez Soto}
\affiliation{Departamento de \'{O}ptica, 
Facultad de Ciencias F\'{\i}sicas, 
Universidad Complutense, 28040 Madrid, Spain}

\author{Jonas S\"{o}derholm}
\affiliation{Department of Electronics, 
Royal Institute of Technology (KTH), Electrum 229, 
SE-164 40 Kista, Sweden}

\date{\today}

\begin{abstract}
We demonstrate a systematic approach to sub-wavelength 
resolution lithographic image formation on films covering areas 
larger than a wavelength squared. For example, it is possible to 
make a lithographic pattern with a feature size resolution of 
$\lambda/[2(N+1)]$ by using a particular $2 M$-photon, 
multi-mode entangled state, where $N \leq M$, and banks of 
birefringent plates. By preparing a statistically mixed such a 
state one can form any pixel pattern on a 
$2^{M-N} (N+1) \times 2^{M-N} (N+1)$ pixel grid 
occupying a square with side $L=2^{M-N-1}\lambda$. 
Hence, there is a trade-off between the exposed area, 
the minimum lithographic feature size resolution, and the 
number of photons used for the exposure. We also show 
that the proposed method will work even under non-ideal 
conditions, albeit with somewhat poorer performance.
\end{abstract}


\maketitle

\section{Introduction}

Classically, to create an optical image, one has to 
modulate a wavefront of an electromagnetic wave in
space. The minimum resolvable feature size of an imaged object corresponds,
roughly speaking, to the minimum modulation period 
allowed, which turns to be of the order of the wavelength 
$\lambda$ of the light used. In fact, the Rayleigh criterion 
establishes that the best resolution that can be achieved 
classically is about $\lambda/2$, which is usually known as 
the diffraction limit.

When the quantum nature of  light is considered, one is
naturally confronted with the role that photon fluctuations
play in setting fundamental performance limits for imaging
systems.  Even if all the \textit{technical} noise sources
are eliminated from the imaging system, the corpuscular nature of the photon induces 
fluctuations, or shot noise, that determine a seemingly 
fundamental spatial resolution, or standard quantum limit, of 
about $\lambda/(2 \sqrt{N})$, where $N$ is the average
number of photons. Sub-wavelength imaging have been used in number of applications~\cite{Dragsten,Flock,Kamimura,Jelles,Denk,Putman}
and a careful analysis shows that indeed the shot-noise sets the resolution limit \cite{Denk,Putman}.

However, the quantum  viewpoint allows for strategies 
that could significantly improve the spatial resolution beyond 
the standard quantum limit. A typical way of reducing 
photon-counting noise is by using multi-mode squeezed 
light~\cite{Kolobov 1,Kolobov 2,Kolobov 3,Kolobov,Fabre,Fabre 2}. This possibility has been 
experimentally demonstrated in other precision 
measurement schemes \cite{Min} and allows one to attain an 
optimum spatial resolution proportional to $\lambda/2 N$, usually 
known as the Heisenberg limit. These sub shot-noise 
imaging systems enable resolving, in principle, arbitrarily small details of 
an object in a diffraction-limited optical system. 

However, writing images imposes even more 
stringent requirements than sub-wavelength resolution of spatial features because in image writing one wants to write small details with \textit{high contrast}. A particular field where circumventing the
classical resolution limit is becoming more and more important
is optical lithography, which is the primary tool for 
writing electronic-circuit patterns. Current production
technologies have tended to use light of shorter
wavelengths to fabricate ever-smaller device features. 

It has been known for some time that entangled
photon pairs can be used to achieve Heisenberg limited resolution of time \cite{Bollinger,Huelga,Buzek} and phase~\cite{Rarity,Margolus,Fonseca,Trifonov}, 
but only very recently it has been proposed the use 
of entanglement to increase high contrast image resolution 
indefinitely~\cite{Boto,Kok,Bjork}. The reason that these 
entangled  quantum states show increased resolution 
can be traced back to the fact that they allow the 
modulation period to be as small as $\lambda/(2N)$, 
and thus they approach Heisenberg-limited resolution. 
The process can be envisioned as the photons clustering 
into a $N$-photon quasi-particle with a linear momentum 
$N$ times large as that of a single photon, and  therefore 
with a shorter de Broglie wavelength~\cite{Jacobson}. 
It is the de Broglie  wavelength that ultimately determines 
the interference  resolution. This has been appreciated for 
a long time in  atomic, molecular and solid-state physics, 
but has only recently been noticed for electromagnetic waves.

In an earlier paper we discussed the use of reciprocal 
binomial states in sub-wavelength resolution 
lithography~\cite{Bjork}. Our method works for even 
number of photons $2N$ and it is especially germane 
to determine the  exposure sequence to generate any 
pixellated pattern on a  $(N+1) \times (N+1)$ grid, occupying 
a square with a half  wavelength long side. An advantage 
with the method is that only one particular entangled state 
needs to be generated: all other necessary states can be 
produced  from the first by means of, e.g., a small bank 
of phase plates with a prescribed birefringence. Unfortunately, 
it is not possible to generate larger patterns since the 
deposition methods proposed hitherto all are periodic with a period 
of half a wavelength~\cite{Boto,Kok,Bjork}. Restricting 
the exposure source of the lithography to four modes, 
with pairwise opposite wavevectors, Fig.~\ref{fig: setup}, 
one can only increase the size of the pattern by some 
factor by sacrificing the pattern resolution by exactly 
the same factor. To be able to adjust the size of the 
deposited pattern \textit{independently} of the resolution 
one must use more modes. In this paper we report a 
systematic multi-mode extension to the method we proposed earlier~\cite{Bjork}.

\section{Sub-wavelength lithography}
\subsection{One dimension, two modes}

Our goal is to establish how to create arbitrary 
two-dimensional  patterns on a squared substrate of 
side $L$.  Suppose we have two counter-propagating 
beams in a direction we shall denote $X$, see Fig. \ref{fig: setup}. The beams 
propagate at angles $\pm \theta$ to the normal of the 
substrate. This substrate is coated with lithographic 
resist (in the following we will refer to the resist as the film) 
and situated in the region where the beams overlap. 
In general, provided the coherence lengths of the 
wave packets are much longer than the side of the film, 
we do not need to take into account the mode shapes
and we will assume that they are plane over the side of the
film.

Following Ref.~\cite{Boto},  the lithographic film absorption 
is modeled by an  $M$-photon absorption process. Thus, the 
absorption process can be modeled by the operator 
$\hat{e}^\dagger{}^M \hat{e}^M$ given by 
\begin{equation}
\hat{e}^\dagger{}^M \hat{e}^M \propto 
\left ( \frac{1}{\sqrt{W}} \sum_i \opa_i^\dagger \right )^M
\left ( \frac{1}{\sqrt{W}}\sum_i \opa_i \right )^M,
\label{eq: Ham}
\end{equation}
where $\opa_i$ is the annihilation operator of mode $i$, and 
$W$ is the number of excited modes impinging on the film. 
In this way, higher-order interference effects are naturally brought 
out.   The photosensitive ``grains" in the film must be much 
smaller than the shortest de Broglie wavelength encountered 
in the exposure process. Therefore, from the point of view of a 
``grain", the photon packets in the respective modes will be 
indistinguishable in spite of their different linear momenta, as 
manifested by (\ref{eq: Ham}). 

Let us now discuss the interference in one dimension between a pair of modes, 
labeled $-1, 1$, propagating in the plane defined by 
the $X$ axis and the film normal. We restrict ourselves to 
consider states that are eigenstates $N_1$ of the total photon number in
the modes $\pm 1$. According to our assumptions 
the deposition rate in the substrate $\Delta_M$ is proportional 
to the expectation value of the operator $\hat{e}^\dagger{}^M \hat{e}^M$, 
where $\hat{e} = (\opa_{-1} + \opa_{1} )/\sqrt{2}$. 
Let us further assume that the two beams impinge at the angles 
$\theta_{\pm 1}=\pm \pi/2$. The beams will hence strike 
the film surface at grazing incidence. Furthermore, we shall assume that the modes 
are prepared in a two-mode reciprocal binomial state of the general form
\begin{equation}
\ket{\psi^{(N_i)}} = \frac{1}{\sqrt{\mathcal{N}_i}}
\sum_{n=0}^{N_i} \sqrt{n!(N_i - n)!} \ 
\ket{n}_i \otimes \ket{N_i - n}_{-i} ,
\label{eq: Relative phase}
\end{equation}
where $N_i$ is the total photon number of the two modes and $\mathcal{N}_i =\sum_{n=0}^{N_i} n!(N_i -n)!$ is a normalization 
factor. 

Let the $X$ coordinate normalized to the optical wavelength $\lambda$ be denoted $x$. Since the  two modes $-1$ and $1$ impinge over the film in anti-parallel 
directions, the accumulated phase of mode $1$ (propagating in the positive 
$X$ direction) at a distance $\lambda x$ from the left edge of the film 
will be 
\begin{equation} 
\hat{U}_1 =\exp(i k \lambda x \hat{a}_1^\dagger \hat{a}_1) = 
\exp(i 2 \pi x \hat{a}_1^\dagger \hat{a}_1) ,
\label{eq: phase}
\end{equation}
where $k=2 \pi/\lambda$, while mode $-1$ will have 
accumulated the phase
\begin{equation} 
\hat{U}_{-1} = \exp[i k \lambda (1-x) 
\hat{a}_{-1}^\dagger \hat{a}_{-1}] = 
\exp[i 2 \pi (1-x) \hat{a}_{-1}^\dagger \hat{a}_{-1} ]  
\label{eq: phase 2}
\end{equation}
at the same location. Using these free-space unitary 
propagation operators, we find that at the location $x$, 
the state (\ref{eq: Relative phase}) for modes $\pm 1$ 
is transformed into 
\begin{equation}
\ket{\psi^{(N_1)}_x} = \frac{1}{\sqrt{\mathcal{N}_1}} 
\sum_{n=0}^{N_1} e^{i  2 \pi x(2 n -N_1)} \sqrt{n!(N_1-n)!} 
\ket{n}_1 \otimes \ket{N_1-n}_{-1} .
\label{eq: Relative phase x}
\end{equation}

We can now translate the substrate a distance $\lambda/[4(N_1+1)]$ 
to the left, and at the same time phase-shift mode $1$ by 
$2 \pi \ell_{1x}/(N_1+1)$ $(\ell_{1x}=1,2,\ldots,N_1+1)$ relative to 
mode $-1$. The corresponding state will be labeled 
$\ket{\psi_x^{(N_1,\ell_{1x})}}$, where
\begin{eqnarray}
\ket{\psi^{(N_1,\ell_{1x})}_x} & = & 
\frac{1}{\sqrt{\mathcal{N}_1}} 
\sum_{n=0}^{N_1} e^{i  \pi(2  x-\frac{\ell_{1x}-1/2}{N_1+1})(2 n -N_1)} 
\nonumber \\
& \times & \sqrt{n!(N_1-n)!}  \
\ket{n}_1 \otimes \ket{N_1-n}_{-1} .
\label{eq: Relative phase x 3}
\end{eqnarray}
As shown in \cite{Bjork}, this state will deposit a \textquotedblleft one-dimensional pixel\textquotedblright, that is, the deposition rate $\Delta_M$ will have a single pronounced peak $\lambda/2(N_1+1)$ wide, occupying the interval on the $X$-axis between $\lambda (\ell_{1x}-1)/2(N_1+1)$ and $\lambda \ell_{1x}/2(N_1+1)$. 

To make a qualitative comparison between the sub-wavelength resolution  
lithographic method proposed in Refs.~\cite{Boto,Kok} and our method, 
we have calculated the deposition pattern when the target pattern 
is a rectangular trench. In Refs.~\cite{Boto,Kok} such a trench, 
$\lambda/4$ wide, was used as a trial target function for a 10-photon state. 
We have done the same, but since a 10-photon state will define a 11-pixel 
pattern (in one dimension) the natural target trench function in our case 
is an integer number of pixel width wide. The pixel width for a 10-photon 
state is $\lambda/22$. In Fig.~\ref{fig: trench} we have calculated the 
deposition rate for a four pixel wide trench, that is, a trench 
$2 \lambda/11 \approx 0.18 \lambda$ wide. In order to make this pattern we can e.g. expose the pixels sequentially employing the states indicated in the figure caption. Although not shown in the figure, remember that this two-mode deposition rate is periodic with the period $\lambda/2$.

We see that the result of our method is almost the same, both in the respect of edge sharpness and in exposure 
penalty, to those obtained by the method proposed by Boto \textit{et al.}~\cite{Boto,Kok}. 
As we shall show below, neither the edge sharpness nor the exposure 
penalty need to be sacrificed when the lithographic pattern is extended 
over areas larger than a half a wavelength in each dimension. A fundamental 
difference between the methods is that the pattern producing state 
is pure in the proposal of Boto \textit{et al.}, while our proposal is 
based on mixed states (or a sequence of pure states if each pixel is deposited separately).

\subsection{One dimension, four modes}

To overcome the limiting $\lambda/2$ periodicity of the deposition rate we introduce another pair of modes $-2$ and $2$ impinging along the $X$ directions at angles $\theta_{\pm 2}=\pm \arcsin[(N_2+1)^{-1}]$ from the film normal. Hence, they have only the wavevector components $\pm 2 \pi/[\lambda(N_2+1)]$ in 
the $X$ direction. We assume that this pair of modes are prepared in a reciprocal binomial state, too. Consequently, their state at location 
$x$ is given by
\begin{equation}
\ket{\phi^{(N_2)}_x} = \frac{1}{\sqrt{\mathcal{N}_2}} 
\sum_{n=0}^{N_2} e^{i  2 \pi x\frac{2 n -N_2}{N_2+1}} 
\sqrt{n!(N_2-n)!}  \ket{n}_2 \otimes \ket{N_2-n}_{-2} .
\label{eq: Relative phase x 2}
\end{equation}
(We use the symbol $\phi$ in the ket above to indicate 
that the modes corresponding to the state do not impinge 
at grazing incidence over the film surface.)

Now suppose the $N_1 + N_2$ photon, product state 
$\ket{\psi^{(N_1)}_x} \otimes \ket{\phi^{(N_2)}_x}$ is 
prepared. Calculating the pattern deposition rate $\Delta_M$, where now $\hat{e} = (\opa_{-2} + \opa_{-1} + \opa_{1} + \opa_{2})/2$ and
$M = N_1 + N_2$, we find that
\begin{eqnarray}
\Delta_M & \propto &
\left  |\sum_{m=0}^{N_1} e^{i  2 \pi x(2 m -N_1)} 
\sum_{n=0}^{N_2} e^{i  2 \pi x\frac{2 n -N_2}{N_2+1}}\right |^2 
\nonumber \\
& \propto &
\frac{1}{[(N_1+1)(N_2+1)]^2} 
\frac{\sin^2 [2 (N_1+1) \pi x]}{\sin^2[2 \pi x/(N_2+1)]}.
\label{eq: Absorption}
\end{eqnarray}
The deposition rate $\Delta_M$ has a highest oscillation period in 
$x$ of $1/[2 (N_1+1)]$ and an overall periodicity of $(N_2+1)/2$, 
corresponding to the physical lengths $\lambda/[2(N_1+1)]$ and 
$\lambda(N_2+1)/2$, respectively. A plot of (\ref{eq: Absorption}) 
for the case $N_1=N_2=3$ is shown in Fig. \ref{fig: raw pixel}. Note that the deposition function spatial resolution is $\lambda/8$ and its periodicity is $2 \lambda$.

When we translate the substrate a distance $\lambda/[4(N_1+1)]$ 
to the left, and at the same time phase-shift mode $2$ by $2 \pi \ell_{2x}/(N_2 +1)+ 
2 \pi \ell_{1x}/[(N_1+1)(N_2+1)]$ $(\ell_{2x}=1,2,\ldots,N_2+1)$ 
relative to mode $-2$, the state $\ket{\phi^{(N_2)}_x}$ in modes $\pm 2$ is transformed to
\begin{eqnarray}
\ket{\phi^{(N_2,\ell_{1x},\ell_{2x})}_x} & = &  
\frac{1}{\sqrt{\mathcal{N}_2}} 
\sum_{n=0}^{N_2} e^{i \pi (2  x-\ell_{2x}-\frac{\ell_{1x}-1/2}{N_1+1})
\frac{2 n -N_2}{N_2+1}} \nonumber \\
& \times & \sqrt{n!(N_2-n)!} \
\ket{n}_2 \otimes \ket{N_2-n}_{-2} .
\label{eq: Relative phase x 4}
\end{eqnarray}

Using (\ref{eq: Relative phase x 3}) and (\ref{eq: Relative phase x 4}) we see that the four-mode state $\ket{\psi^{(N_1)}_x} \otimes \ket{\phi^{(N_2)}_x}$ will consequently be transformed into the state $\ket{\psi_x^{(N_1,\ell_{1x})}} \otimes 
\ket{\phi_x^{(N_2, \ell_{1x},\ell_{2x})}}$ after the translation and respective relative phase shifts. The deposition rate for this state can  readily be calculated to be
\begin{eqnarray}
\Delta_M(\ell_{1x},\ell_{2x}) & \propto &
\frac{1}{[(N_1+1)(N_2+1)]^2} \nonumber \\
&\times &  \frac{\sin^2\{[2 (N_1+1)x-\ell_{1x}+1/2]\pi \}}
{\sin^2 \left [ \left (2 x- \ell_{2x} -\frac{\ell_{1x}-1/2}{N_1+1} \right ) 
\frac{\pi}{N_2+1} \right ]} .
\label{eq: Absorption 2}
\end{eqnarray}
If we divide the part of the $X$ axis between the 
origin and the point $x=(N_2+1)/2$ into $(N_1+1)(N_2+1)$ pieces, each $1/[2(N_1+1)]$ long, each interval will represent a \textquotedblleft one-dimensional pixel\textquotedblright. With a 
specific choice of $\ell_{1x}$ and $\ell_{2x}$ we can deposit 
(or expose) pixel number $\ell_{1x}+(N+1)\ell_{2x}$ [numbered 
from left to right and the number taken modulo $(N_1+1)(N_2+1)$] 
with a negligible exposure penalty (that is, negligible unwanted exposure of nominally unexposed pixels). This can be 
clearly seen in Fig.~\ref{fig:four modes}, where we have assumed 
that  $N_1=N_2=3$, $\ell_{1x}=2$, and $\ell_{2x}=1$, leading to 
the exposure of pixel number 6. The relative phase-shifts between 
the modes, labeled by the numbers $\ell_{1x}$ and $\ell_{2x}$, can 
be accomplished via a bank of appropriately chosen birefringent plates, 
provided that the modes $\pm 1$, and $\pm 2$, respectively, are 
originally prepared in spatially and temporally degenerate modes, but with 
orthogonal polarizations as discussed in Ref.~\cite{Bjork}. 

The deposition-rate function (\ref{eq: Absorption 2}) has two special 
properties that are worth pointing out. The first is that the deposition rate will 
be identically zero at the center of all unexposed pixels regardless 
of what other pixels are exposed. Hence, a nominally unexposed pixel 
surrounded by exposed pixels will remain unexposed at the pixel center. 
This is a very appealing feature of the proposed method since the exposure 
penalty hardly depends at all on the particular pixel pattern one intends to 
expose. The second nice feature is that the sum of the deposition-rate 
functions for all pixels add up to unity; i.e.
\begin{equation}
\sum_{\ell_{1x}=1}^{N_1+1}\sum_{\ell_{2x}=1}^{N_2+1}
\Delta_M(\ell_{1x},\ell_{2x}) \equiv 1  ,
\label{eq: Absorption 3}
\end{equation}
for all values of $x$. This, in turn, means that we never risk overexposure, even if we expose two 
or more adjacent pixels. In fact, if a row, or column, of adjacent 
pixels are exposed, the resulting deposition function ridge will hardly 
have any modulation~\cite{Bjork}. The identity~(\ref{eq: Absorption 3}) 
also means that if one wants to make the negative image of some pixel 
pattern one can construct \textit{identically} the negative image deposition rate 
by exposing all previously unexposed pixels, and vice versa.

Let us now discuss the geometrical scaling properties of 
the deposition rate. By decreasing the modes' wavevector components in the film plane by a 
fixed factor, both the minimum feature size resolution and the 
fundamental period of deposition rate will increase by the same 
factor. If we, e.g., let modes $\pm 1$ impinge at angles 
$\pm \arcsin(1/2)$ from the film normal and modes $\pm 2$ impinge 
at the angles $\pm \arcsin([2(N_2+1)]^{-1})$, then the minimum 
feature size resolution (i.e., pixel size) becomes $\lambda/(N_1+1)$ 
and the period of deposition rate becomes $(N_2+1) \lambda$. However, 
the wavevector component parallel to the film is not only governed by the modes' propagation angles but are also governed by the de Broglie wavelength of the impinging states. Therefore, 
the pixel and pattern sizes are intimately connected to how we 
prepare the states. If the (one-dimensional) film is modeled as a 
$M=2 N$-photon absorber, the choice to partition the $2 N$ photons equally between the two 
pairs of modes $\pm 1$, and $\pm 2$, as assumed in Fig.~\ref{fig:four modes}, is by no means necessary. Instead we can, e. g., prepare modes $-1$ and $1$ in a two-mode 
$(N-1)$-photon state $\ket{\psi^{(N-1)}_x}$ and the modes $-2$ 
and $2$ in a two-mode $(N+1)$-photon state $\ket{\phi^{(N+1)}_x}$. 
The appropriate relative phase shifts are $2 \pi \ell_{1x}/N$, where 
$\ell_{1x}=1,2, \ldots, N$ and $2 \pi \ell_{2x}/(N +2)+ 
2 \pi \ell_{1x}/(N^2 + 2 N)$ where $\ell_{2x}=1, 2, \ldots, N+2$, 
respectively. In this case, the minimum lithographic feature size 
resolution becomes $\lambda/(2N)$, the number of individually 
depositable pixels become $N(N+2)$, and the fundamental period 
of the deposition rate becomes $\lambda (N+2) / 2$. An illustration 
of an ensuing deposition rate function is given in Fig.~\ref{fig: unbalanced pattern}. 
Continuing this re-partition one can either increase the fundamental 
period of the deposition rate at the expense of increasing the 
minimum resolution by increasing the photon number in modes 
$\pm 2$ at the expense of the photon number in modes $\pm 1$, 
or vice versa. The attainable minimum size resolution and deposition 
rate period are shown in Table~\ref{table: 1}.

\subsection{Two dimensions, eight modes}

One can now extend the lithographic exposure procedure to two 
dimensions by simply introducing two additional pairs of modes 
$\pm 1^\prime$ and $\pm 2^\prime$, impinging towards the film 
at corresponding angles to modes $\pm 1$ and $\pm 2$, but in the 
$Y$ direction, perpendicular to $X$~\cite{Bjork}. If the eight modes 
are prepared in the initial state $\ket{\psi_x^{(N,\ell_{1x})}} \otimes 
\ket{\phi_x^{(N, \ell_{1x},\ell_{2x})}} \otimes
\ket{\psi_y^{(N,\ell_{1y})}} \otimes 
\ket{\phi_y^{(N, \ell_{1y},\ell_{2y})}}$, the deposition rate is given 
by the product of the corresponding deposition rates in the $X$ and in 
the $Y$ directions. Of course, if the number of photons in each of the 
two-mode states $\ket{\psi_x^{(N,\ell_{1x})}}$, \ldots, 
$\ket{\phi_y^{(N, \ell_{1y},\ell_{2y})}}$ is $N$, then the film must 
have a non-negligible $M=4 N$-photon absorption cross section. If so, 
the assumed state will expose the pixel $(\ell_{1x}+(N+1)\ell_{2x},
\ell_{1y}+(N+1)\ell_{2y})$ and leave the remaining pixels essentially 
unexposed. In order to expose a pattern, such as a line of adjacent 
pixels, one must prepare a statistical mixture of the pixels' associated 
states or one can expose the pixels sequentially. 
As shown in Ref.~\cite{Bjork}, diagonal lines composed of 
exposed pixels will have an unacceptably large deposition rate 
fluctuation along the diagonal center-line. However, the minima 
can be ``filled in"  by depositing intermediate pixels (with their centers 
at the intersection points between four adjacent regular pixels). 
Again, the states corresponding to these intermediate pixels can be 
prepared by appropriate relative phase shifts between the mode 
pairs~\cite{Bjork}. For the rest of this paper we shall only study 
the deposition rates in one dimension, bearing in mind that with 
our method the two-dimensional deposition rate function is simply 
the product of two one-dimensional functions.

\section{Generalized multi-mode quantum lithography}

It is clear that the procedure to increase the deposition rate 
period is not limited by considering only two pairs of modes 
with opposite wavevectors in each dimension. We can continue this procedure 
by introducing a third pair of modes, labeled $-3$ and $3$, impinging 
towards the film at the angles $\theta_{\pm 3}=\pm \arcsin[1/(N+1)^2]$ 
from the film normal. If the number of photons $M$ contributing 
to the film absorption process is divisible by 3 (or 6, in two 
dimensions), so that $M=3 N$, and this photon number is 
partitioned equally between the three modes, then one will 
be able to deposit any pixellated patterns with the minimum 
feature size resolution of $\lambda/[2(N+1)]$ over a length of 
$L=\lambda (N+1)^2/2$. 

However, in order to cover the maximum area for a given 
number of photons $M$ and resolvable feature size 
$\lambda/[2 (N+1)]$, where $1 \le N < M$, the following 
product state should be prepared:
\begin{equation} 
\ket{\psi_x^{(N,\ell_{1x})}} \otimes 
\ket{\phi_x^{(1,\ell_{1x},\ell_{2x})}} \otimes \ldots 
\otimes \ket{\phi_x^{(1,\ell_{1x},\ell_{2x}, \ldots, \ell_{(M-N)x})}},
\label{eq: optimal state}
\end{equation}
where $\ell_{1x}=1,2, \ldots, N+1$, $\ell_{2x}, \ldots, 
\ell_{(M-N)x}=1,2$, and modes $\pm 1$ impinge at grazing 
incidence while modes $\pm i$, $i=2, \ldots , M-N$, impinge 
at the angles $\theta_{\pm i}=\pm \arcsin(2^{-(i-1)})$. 
The rationale for preparing this state is that the state 
$\ket{\psi_x^{(N,\ell_{1x})}}$ will determine the feature 
size resolution and let us deposit any one of $N+1$ pixels 
each with a size of $\lambda/[2(N+1)]$. With the remaining 
$M-N$ photons, each of the one photon states 
$\ket{\phi_x^{(1,\ell_{1x},\ell_{2x})}}, \ldots , 
\ket{\phi_x^{(1,\ell_{1x},\ell_{2x}, \ldots, \ell_{(M-N)x})}}$ 
will allow us to double the fundamental period of the deposition 
rate function. If we compare this to a case where the $M-N$ 
photons are partitioned between a smaller number of more 
highly excited states, it is clear that the state~(\ref{eq: optimal state}) 
gives a longer fundamental deposition rate period since 
$2^{M-N}\geq M-N+1$ for all relevant numbers $M$ and $N$. 
With the initial state~(\ref{eq: optimal state}) one will be able 
to deposit any one of $2^{M-N} (N+1)$ pixels in one 
dimension, where each pixel is $\lambda/[2(N+1)]$ wide. 
The fundamental period of the pixel pattern will be 
$L=2^{M-N-1}\lambda$. With $2 M$ photons one will 
be able to make a two-dimensional pattern with this resolution 
and periodicity in both dimensions. This is the major result in 
this paper. In Fig.~\ref{fig: eight modes} we have plotted 
the one-dimensional deposition rate from the interference 
between eight modes in a statistical mixture of the states
\begin{equation} 
\ket{\psi_x^{(4,1)}} \otimes 
\ket{\phi_x^{(1,1,1)}} \otimes 
\ket{\phi_x^{(1,1,1,1)}} \otimes 
\ket{\phi_x^{(1,1,1,1,0}}
\label{eq: optimal state 2}
\end{equation}
and
\begin{equation} 
\ket{\psi_x^{(4,3)}} \otimes 
\ket{\phi_x^{(1,3,1)}} \otimes 
\ket{\phi_x^{(1,3,1,1)}} \otimes 
\ket{\phi_x^{(1,3,1,1,0}},
\label{eq: optimal state 3}
\end{equation}
where modes $\pm 1 $ impinge at $\theta_{\pm 1}=\pm \pi/2$ 
and the remaining three pairs of modes impinge at the angles 
$\theta_{\pm 2}= \pm \arcsin(1/2)$, $\theta_{\pm 3}= \arcsin(1/4)$, 
and $\theta_{\pm 4}=\pm \arcsin(1/8)$. The total photon number 
in both states is six, the same number assumed in 
Fig.~\ref{fig:four modes} and Fig.~\ref{fig: unbalanced pattern}. 
The ensuing 32-pixel pattern is the largest one-dimensional 
pattern one can make with six photons provided that the minimum 
feature size is fixed to $\lambda/8$. The price for the large number 
of depositable pixels is the difficulty one will have generating this 
eight-mode state. If the states' relative phase-shifts are generated 
by birefringent phase plates, one will need two plates for the first 
pair of modes, and three, four and five plates for the remaining 
three pairs of modes, respectively. We can in principle continue 
this process \textit{ad infinitum}, but for a fixed minimum feature 
size, this requires $M$, the number of photons contributing to 
the exposure process, to increase. Since the absorption cross 
section quite generally decreases rapidly with increasing $M$, 
there will be a practical limit to such an extension.

\section{Imperfections due to losses and competing multi-photon 
absorption processes}

Above we have discussed  how sub-wavelength imaging can 
work under ideal conditions. However, in order for the proposed 
method to be of practical use it is necessary that it is robust 
against imperfections. Below we shall discuss three mechanisms 
that will deteriorate the image forming ability of entangled states: 
linear losses,  competing multi-photon absorption processes, and 
exposure noise due to light quantization.

First we will discuss the effect of losses occuring between the 
state generator and the film. As long as the film strictly absorbs 
the same number of photons as the generated multi-mode state contains, 
losses will not affect the lithographic resolution, it will only lower the 
deposition rate by a fixed amount. This is rather obvious, because 
if one or more photons are lost from an $M$-photon state, no 
$M$-photon absorption process can be triggered by the state. 
However, if there exists a competing $(M-1)$-photon absorption 
process in the film, the film may be exposed even after a photon 
is lost. In this case the modified deposition rate will be the same 
whether the photon is lost before impinging on the film or if only 
$M-1$ out of $M$ photons impinging on the film are absorbed. Losses prior 
to the film will, however, shift the relative proportion between $(M-1)$- 
and $M$-photon absorption processes in favor of the former 
by decreasing the probability that the state impinging on the film 
contains $M$ photons. Therefore, losses in the optical system 
prior to the film should be kept as low as possible. Fortunately, 
losses will only gradually increase the required exposure dose 
and shift the probability of absorption toward absorption 
processes involving a smaller number of photons.

Next, we shall examine how the deposition rate of a $M$-photon 
state is affected by absorption processes involving less than $M$ 
photons. Two physical effects will deteriorate the resolution and 
the exposure penalty in this case. One is that if the state contains 
$M$ photons, but only $M-1$ of them are absorbed, there are as 
many final states as there are modes. If we look at the simplest 
case, an impinging two-mode state $\ket{\psi^{(M)}_x}$, the 
possible final states are $\ket{1,0}$ and $\ket{0,1}$. Since 
these states are distinguishable, the absorption probability amplitudes 
leading to one of these final states will not interfere with 
the amplitudes leading to the other. Only the initial state 
$\ket{M,0}$ ($\ket{0,M}$) will evolve into the state 
$\ket{1,0}$ ($\ket{0,1}$) with certainty upon absorption of 
$M-1$ photons. (All other number-difference states 
$\ket{M-n,n}$, $n \neq 0,M$ can evolve either into 
$\ket{1,0}$ or $\ket{0,1}$). This means that the two extreme 
number-difference states in the expansion of $\ket{\phi^{(M)}_x}$ 
cannot interfere in a $M-1$ photon absorption process. Therefore, 
the Fourier component with the highest spatial frequency will be 
absent in the ensuing deposition rate. Hence, the \textit{spatial 
resolution} will decrease monotonically with decreasing order 
of the absorption process. A second effect will be that the destructive interference 
between the absorption probabilities outside the designated pixel 
will be incomplete, leading to an increased \textit{exposure penalty}. 

In Fig. \ref{fig: smaller absorption} we have drawn the 
deposition rates of the state $\ket{\psi_x^{(4,3)}}$ due to 
4-, 3-, 2-, and 1-photon absorption processes. The curves have been 
normalized such that the deposition rates all have a maximum 
of unity to facilitate comparisons. It can be seen that the 
width of the deposition rate peak increases, and so does 
the exposure penalty as the mismatch between the state 
and the absorption process increases. It is, of course, 
possible to make a better deposition rate function 
for, e. g., a 3-photon absorption film by exposing the film 
by a state of the kind $\ket{\psi^{(3)}_x}$ instead of the 
state $\ket{\psi^{(4,3)}_x}$. On the positive side, it is 
seen that if the film allows both for a 3-photon and a 4-photon
absorption process, our method will yield a somewhat poorer 
result than if the 3-photon cross-section were identically zero. 
Still, the ensuing deposition rate is moderately deteriorated 
as compared to the ideal case.

In Fig. \ref{fig: smaller absorption 2} we have drawn similar 
curves for a six-mode, four photon state that exposes pixel 7 
(from the left) of 12 pixels. Each pixel is $\lambda/6$ wide, 
and the fundamental period of the deposition rate is $2 \lambda$. 
Again, 3-, 2-, and 1-photon absorption processes will deteriorate 
the deposition rate, and in this case by a larger amount 
than for the two-mode state. The physical reason is that 
with a larger number of modes, there can also be a larger 
number of final states that will separate the interference 
paths into a larger number of distinguishable groups. This 
will primarily affect the (destructive) interference between 
the different absorption paths outside the deposition rate peak. 
However, in the six-mode case too, the deterioration is gradual. 
Hence, even if the 3-photon cross section is not identically zero, 
the consequences are not catastrophic.

The final effect we wish to discuss regards the fact that the 
calculated deposition rate is an ensemble average. In an 
experiment, the actual deposition rate may look rather 
different than its expectation value. This effect has not 
been discussed in any of the previous papers on 
entangled-state sub-wavelength lithography~\cite{Boto,Kok,Bjork}. 
If a state of the type given by~(\ref{eq: optimal state}) 
impinges on the film, the probability of state absorption is 
with all likelihood low. In addition, as have been discussed 
previously, each pixel defined on the film must contain many 
photo-sensitive grains since, for the method to make sense,
 the grains must be smaller (preferably much smaller) than 
the pixel size. Therefore, in order to expose a pixel, many 
exposure shots (per pixel) are needed. Since the absorption 
process is stochastic, the ensuing exposure will also be 
stochastic. The consequences of such an effect was 
studied in the context of the opposite process, namely image 
recognition, by Rose already in the 1940s~\cite{Rose,Rose 2} 
(an up-to-date review of this early work was recently 
written by Burgess~\cite{Burgess}). In a first order 
approximation, the exposure of each pixel can be 
modeled by a Poisson distribution with a mean determined 
by the relation between the absorption probability of a grain, 
the deposition rate at the pixel, and the exposure dose. It is 
clear that in order to have pixels with a relative variation in 
exposure of, say less than 10\%, the mean number of 
exposed grains must be larger than 100. This in turn implies 
that the number of states that need to impinge on the film 
to expose this particular pixel must be much larger than 100. 
As a consequence, it is desirable that the state generator 
emits states with a high repetition frequency in order to 
expose the film swiftly.

\section{Conclusions}

We have developed a multi-mode extension of the sub-wavelength lithographic method based on multi-photon absorption proposed in
Ref.~\cite{Bjork}. We have shown the optimal way (in terms of 
the number of photons used in the process) to generate a pixellated 
pattern of given (sub-wavelength) resolution and pattern size. A 
salient feature of our proposal is that only one particular multi-mode 
photon number eigenstate needs to be prepared. Any pixellated 
pattern can be generated from this state by applying differential 
phase-shifts between the modes. For a given lithographic feature 
resolution, the price to be paid for a larger pattern is that the 
complexity of the state used to expose the pattern increases. 
For each doubling of the pattern area, two more modes in a 
one-photon state must be used. This also requires the film 
absorption process to increase by one in terms of photons 
absorbed in the process. Therefore it seems unlikely that it 
will be possible to make very large patterns. We have also 
studied the effects of imperfections, in terms of losses, 
competing lower order multi-photon processes, and 
deposition-rate fluctuations due to the quantization of the 
light used for the exposure. We have shown that neither 
effect results in catastrophic consequences, demonstrating that 
the proposed method is somewhat robust against imperfections.

\begin{acknowledgments}
This work was supported by the Swedish Research Council 
for Engineering Sciences (TFR), the Swedish Foundation for 
Strategic Research (SSF) and L M Ericssons 
stiftelse f\"{o}r fr\"{a}mjande av elektroteknisk forskning.
\end{acknowledgments}


\newpage

\begin{figure}
\caption{A schematic showing the geometry of the lithographic exposure.}
\label{fig: setup}
\end{figure}

\begin{figure}
\caption{The deposition rate due to a pair of modes with $\theta_{\pm 1}=\pm \pi/2$ and $N=10$ in an equal statistical mixture state between $\ket{\psi_x^{(10,1)}}$, $\ket{\psi_x^{(10,2)}}$, $\ket{\psi_x^{(10,3)}}$, $\ket{\psi_x^{(10,4)}}$, $\ket{\psi_x^{(10,9)}}$, $\ket{\psi_x^{(10,10)}}$, and $\ket{\psi_x^{(10,11)}}$. A trench, 4 pixels wide ($= 4 \lambda/22$) is formed. The unwanted exposure modulation of both the exposed and the unexposed pixels is on the order of 10\% of the maximum exposure. The target deposition function is drawn with thick lines.}
\label{fig: trench}
\end{figure}

\begin{figure}
\caption{The deposition rate due to two pairs of modes with $\theta_{\pm 1}=\pm \pi/2$ and $\theta_{\pm 2}=\pm \arcsin(1/4)$ that are prepared in two three-photon reciprocal binomial states. The deposition rate is periodic with the period $2 \lambda$. The width of the deposition rate peak is roughly $\lambda/8$.}
\label{fig: raw pixel}
\end{figure}

\begin{figure}
\caption{The deposition rate due to two pairs of modes with $\theta_{\pm 1}=\pm \pi/2$ and $\theta_{\pm 2}=\pm \arcsin(1/4)$ in the six photon reciprocal binomial product state $\ket{\psi_x^{(3,2)}} \otimes \ket{\phi_x^{(3, 2,1)}}$ exposing pixel number six. The deposition rate is periodic with the period $2 \lambda$. Only one period is shown.}
\label{fig:four modes}
\end{figure}

\begin{figure}
\caption{The deposition rate due to two pairs of modes with $\theta_{\pm 1}=\pm \pi/2$ and $\theta_{\pm 2}=\pm \arcsin(1/5)$ in the state $\ket{\psi_x^{(2,1)}} \otimes \ket{\phi_x^{(4, 1,1)}}$. In comparison to the example in Fig. \ref{fig:four modes} the pixel size is increased by a factor $4/3$ to $\lambda/6$ while the deposition rate period has increased by a factor 5/4 to $5 \lambda/2$.}
\label{fig: unbalanced pattern}
\end{figure}

\begin{figure}
\caption{The deposition rate due to a superposition state of four pairs of modes with three, one, one and one photon. The pixel size is $\lambda/8$ while the deposition rate period has increased to $4 \lambda$. The relative phase shifts of the states have been chosen so that pixels 13 and 15 are exposed. As can be seen, the deposition rate is zero at the center of pixel 14.}
\label{fig: eight modes}
\end{figure}

\begin{figure}
\caption{The deposition rate due to a pair of modes with $\theta_{\pm 1}=\pm \pi/2$ in the state $\ket{\psi_x^{(4,3)}}$. The solid line shows the deposition rate for a 4-photon absorption film. The dashed, dotted, and dash-dotted line represent the (normalized) deposition rate for 3-, 2-, and 1-photon absorption processes, respectively.}
\label{fig: smaller absorption}
\end{figure}

\begin{figure}
\caption{The deposition rate due to three pairs of modes with $\theta_{\pm 1}=\pm \pi/2$, $\theta_{\pm 2}=\pm \arcsin(1/2)$, and $\theta_{\pm 3}=\pm \arcsin(1/4)$ in the four photon state $\ket{\psi_x^{(2,1)}} \otimes \ket{\phi_x^{(1,1,2)}}\otimes \ket{\phi_x^{(1,1,2,2)}}$. The solid line shows the deposition rate for a 4-photon absorption film. The dashed, dotted, and dash-dotted line represent (normalized) the deposition rate for 3-, 2-, and 1-photon absorption processes, respectively.}
\label{fig: smaller absorption 2}
\end{figure}

\begin{table}
\caption{A table demonstrating the number of depositable pixels, minimum feature size resolution, and the fundamental deposition rate periodicity for different partitions of $N_1 + N_2 =2 N$ photons between two pairs of modes in one dimension.}
\begin{tabular}{ccccc}
$N_1$&$N_2$&Number of pixels&Feature size&Periodicity\\
\colrule
$2 N$&$0$&$2N+1$&$\lambda/(4N+2)$&$\lambda/2$\\
$2 N-1$&$1$&$4N$&$\lambda/4N$&$\lambda$\\
$\vdots$&$\vdots$&$\vdots$&$\vdots$&$\vdots$\\
$n$&$2N-n$&$(n+1)(2N-n+1)$&$\lambda/2(n+1)$&$(2N-n+1)\lambda/2$\\
$\vdots$&$\vdots$&$\vdots$&$\vdots$&$\vdots$\\
$1$&$2N-1$&$4N$&$\lambda/4$&$N\lambda$\\
$0$&$2N$&$2N+1$&$\lambda/2$&$(2N+1)\lambda/2$\\
\end{tabular}
\label{table: 1}
\end{table}

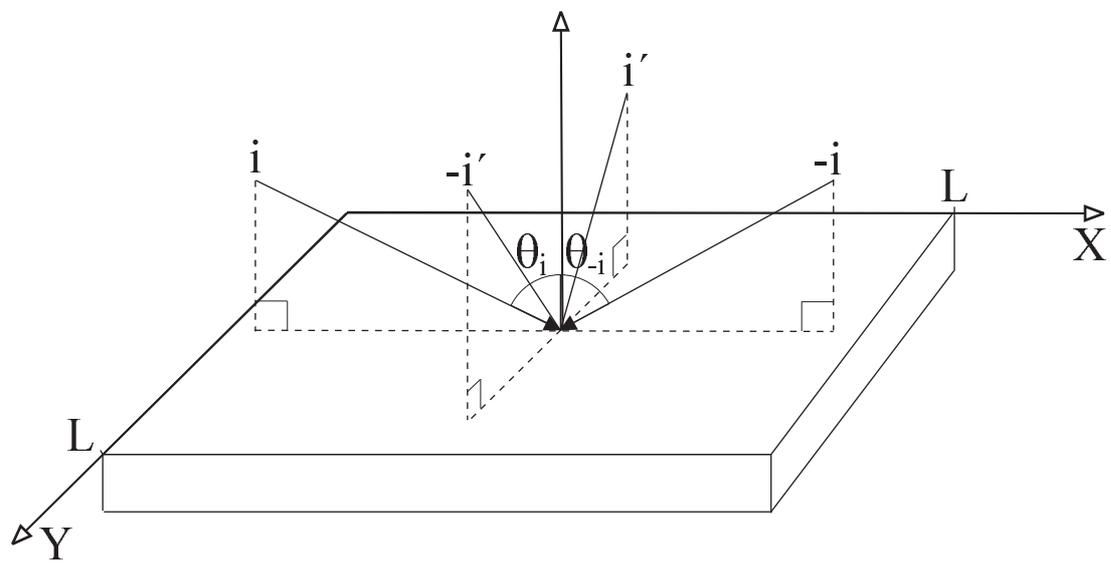
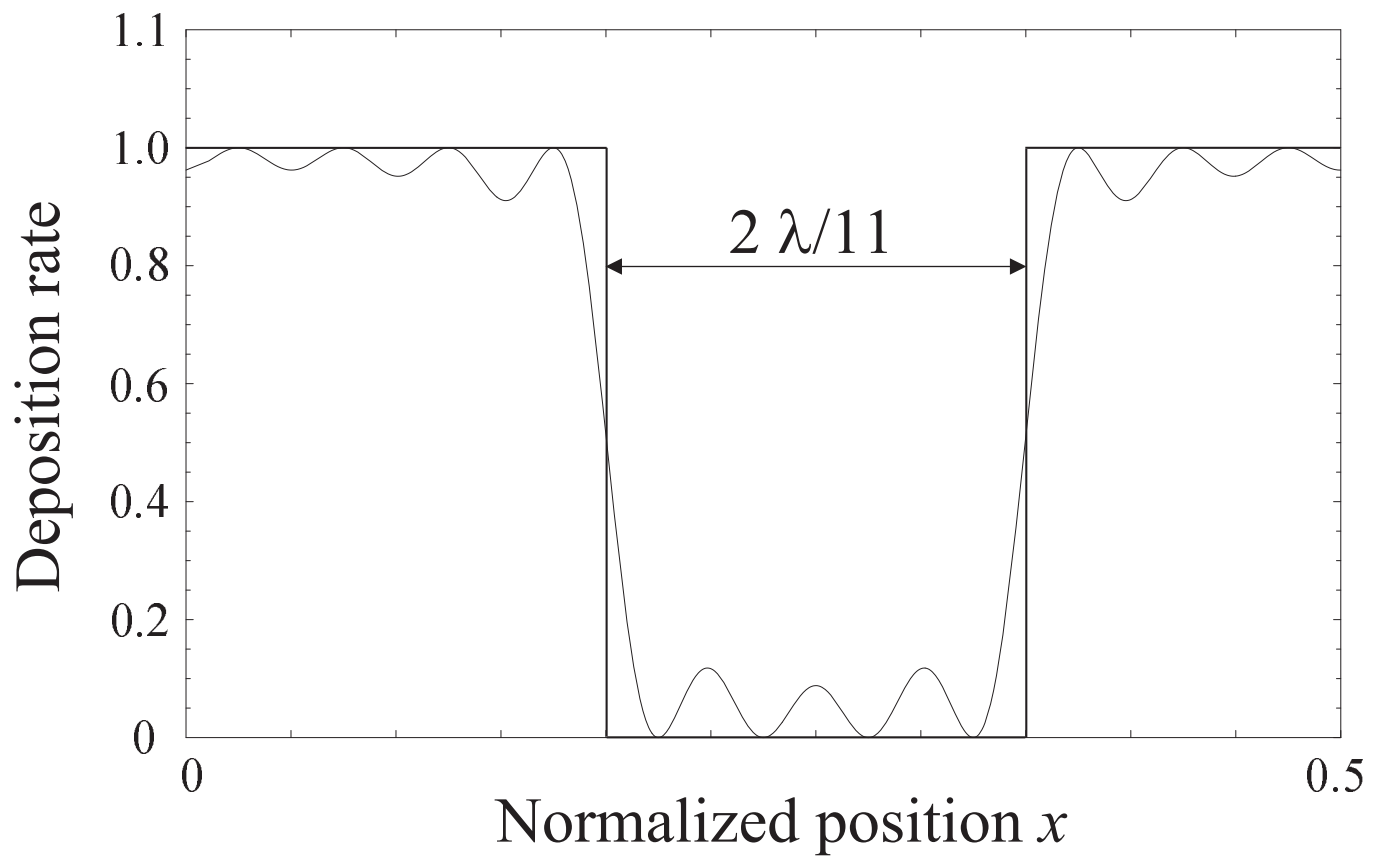
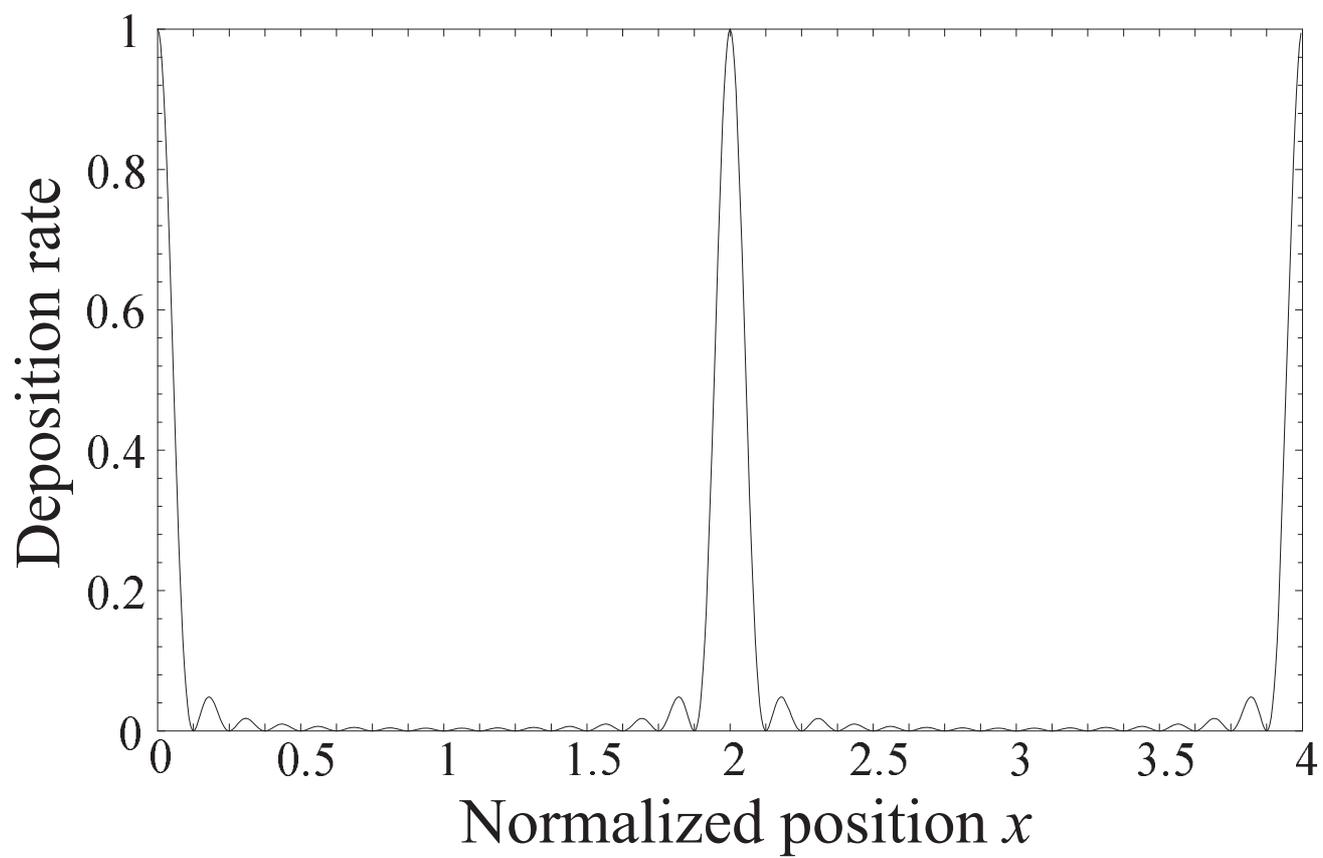
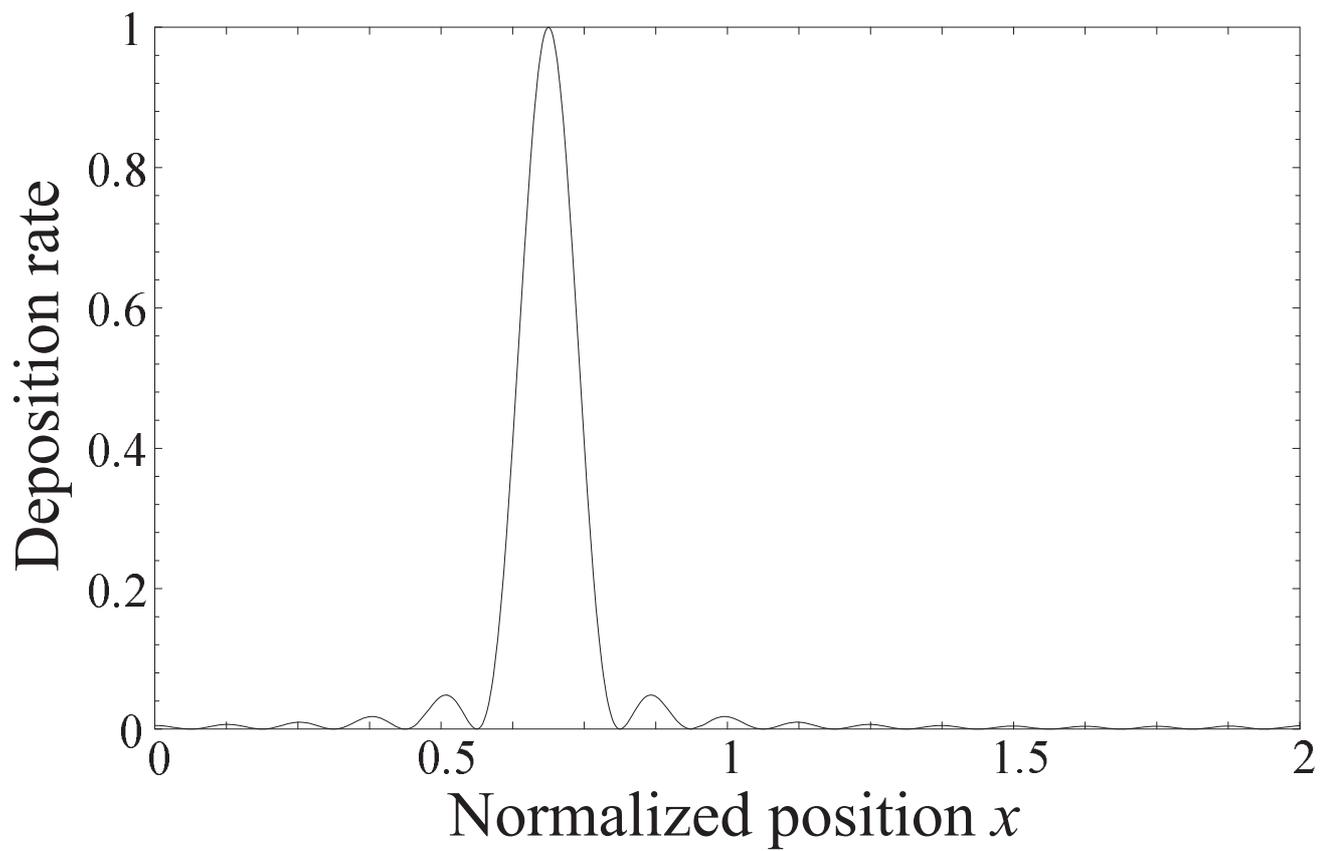
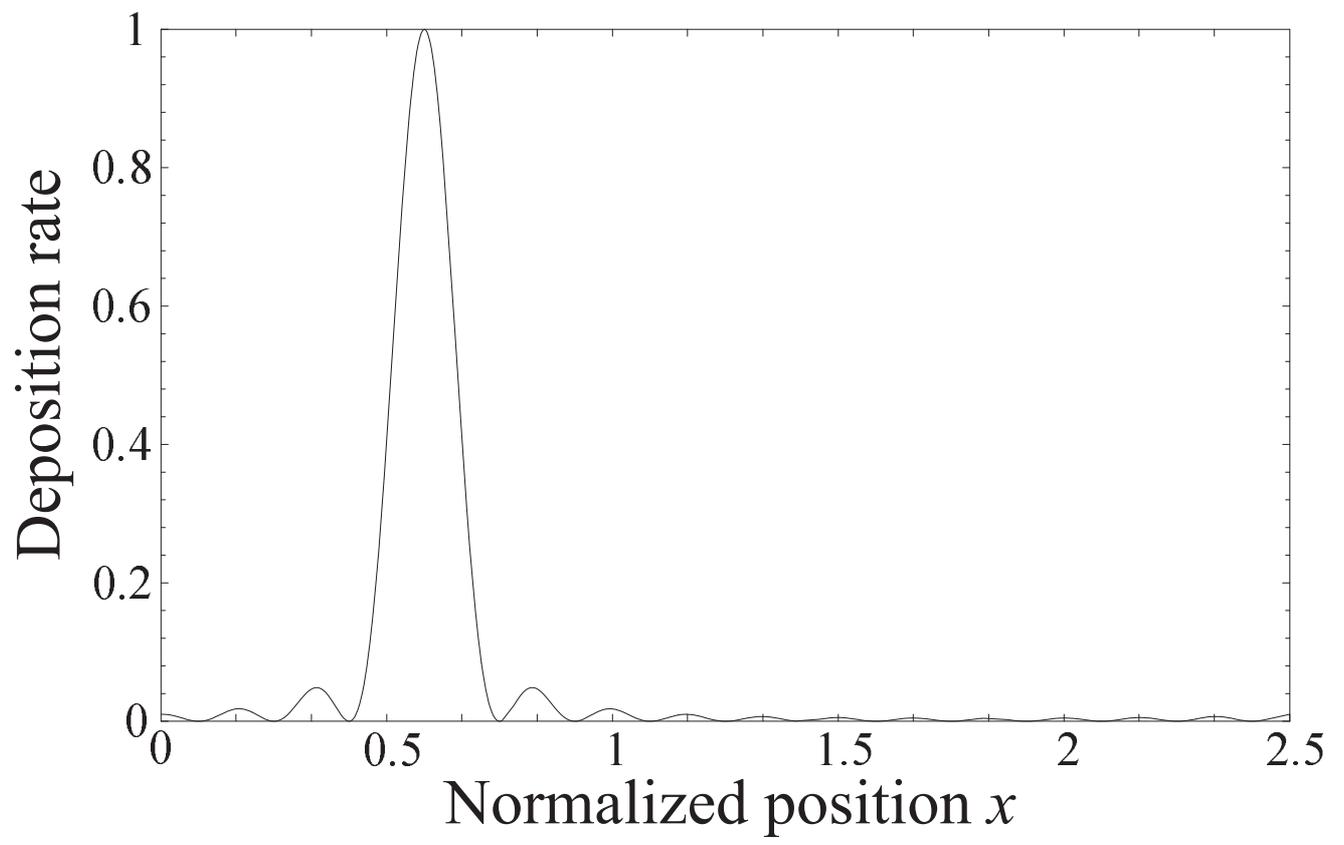
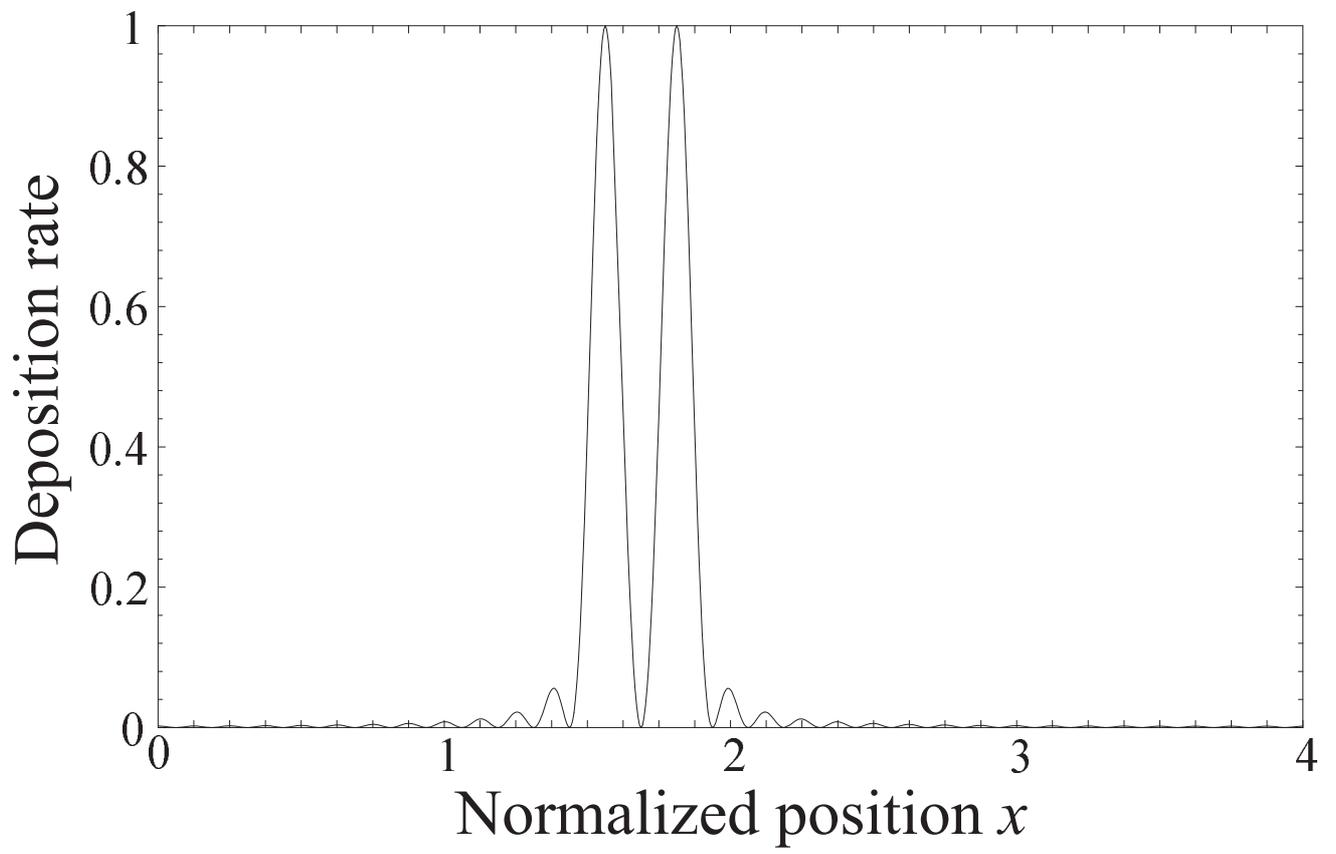
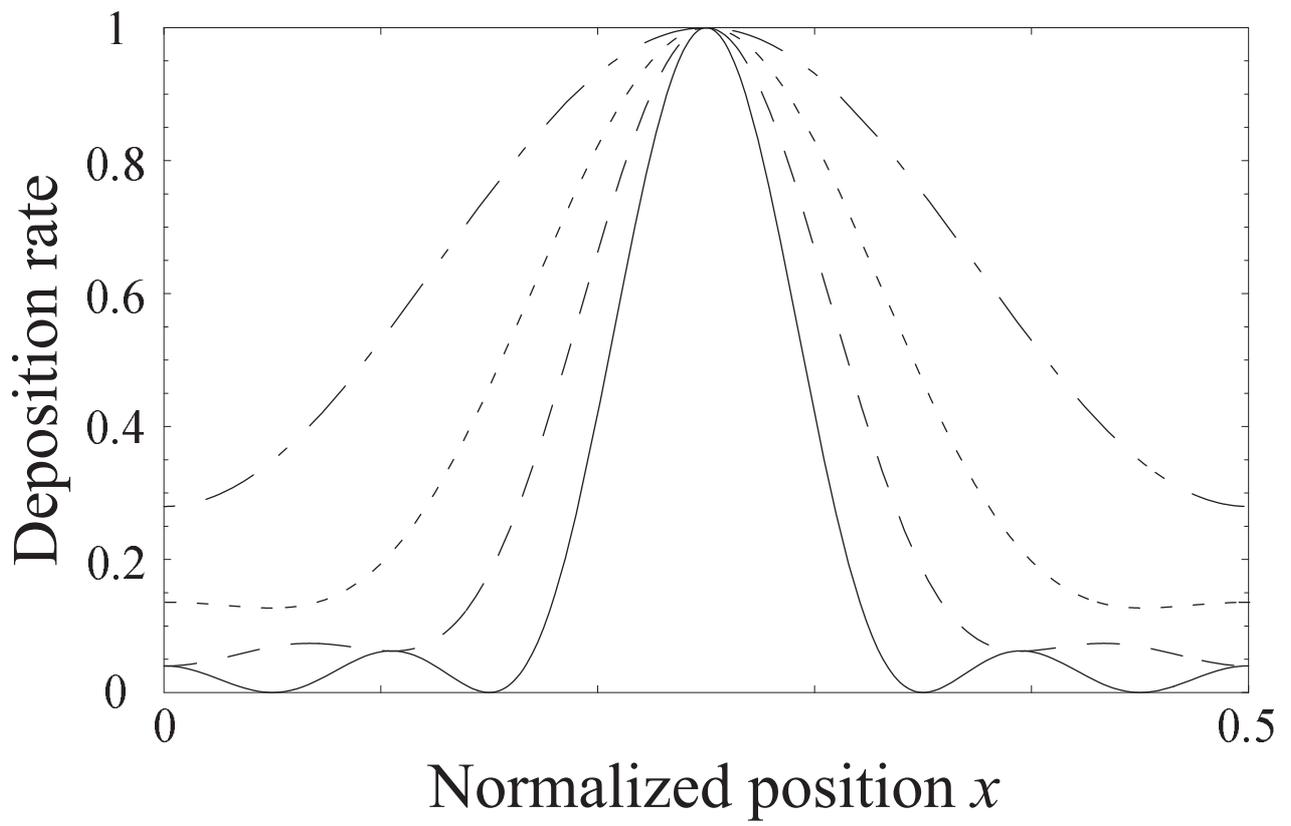
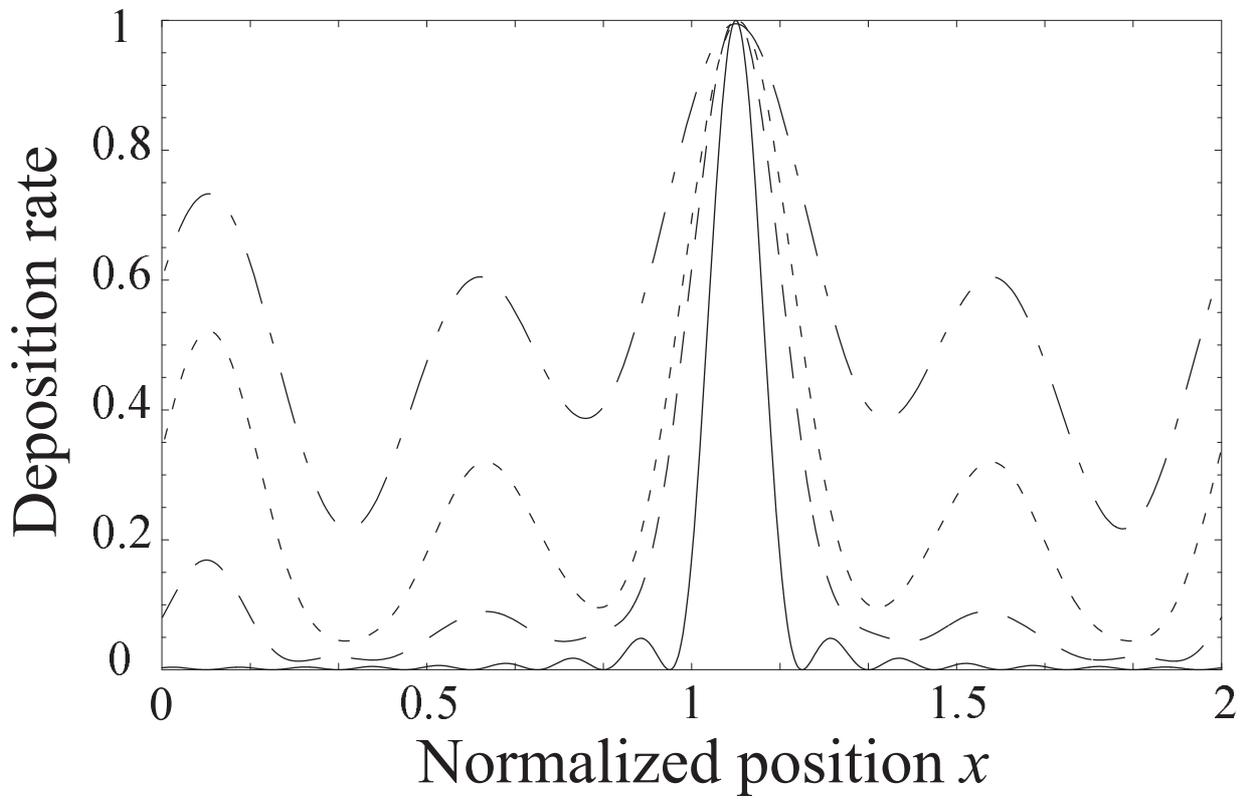
\end{document}